\documentclass[11pt, a4paper]{article}
\usepackage[utf8]{inputenc}
\usepackage[T1]{fontenc}
\usepackage[margin=1in,footskip=0.5in]{geometry}
\usepackage{amsmath}
\usepackage{amssymb}
\usepackage{hyperref}
\usepackage{graphicx}
\usepackage{booktabs,multirow,color}
\usepackage{lscape}
\usepackage{setspace}
\usepackage[round]{natbib}
\numberwithin{equation}{section}

\begin{document}
	
	\begin{center}
		{\Large\bf Bayesian estimation of Unit-Weibull distribution based on dual generalized order statistics with application to the Cotton Production Data\\}
		\vspace{0.5cm}
		Qazi J. Azhad$^a$, Abdul Nasir Khan$^{b}$\footnote{Corresponding Author: A. N. Khan (nasirgd4931@gmail.com )}, Bhagwati Devi$^c$, Jahangir Sabbir Khan$^d$, Ayush Tripathi$^e$\\
		$^a$Department of Mathematics, Shiv Nadar Institution of Eminence, Dadri, India\\
		$^b$Department of Mathematics and Statistics, Dr. Vishwanath Karad MIT World Peace
		University, Pune, India\\
		$^c$Department of Statistics, Central University of Jharkhand, Jharkhand, India\\
	    $^d$Department of Statistics and Operations Research, Aligarh Muslim university, India\\
		$^e$Department of Mathematics, Jaypee Institute of Information Technology, , India\\
		\vspace{0.5cm}
	\end{center}
	\hrule
	\begin{abstract}
		The Unit Weibull distribution with parameters $\alpha$ and $\beta$ is considered to study in the context of dual generalized order statistics. For the analysis purpose, Bayes estimators based on symmetric and asymmetric loss functions are obtained. The methods which are utilized for Bayesian estimation are approximation and simulation tools such as Lindley, Tierney-Kadane and Markov chain Monte Carlo methods. The authors have considered squared error loss function as symmetric and LINEX and general entropy loss function as asymmetric loss functions. After presenting the mathematical results, a simulation study is conducted to exhibit the performances of various derived estimators. As this study is considered for the	dual generalized order statistics that is unification of models based distinct ordered random variable such as order statistics, record values, etc. This provides flexibility in our results and in continuation of this, the cotton production data of USA is analyzed for both submodels of ordered random variables: order statistics and record values.\\
		
		
	\end{abstract}
\section{Introduction}\label{secintro}
\noindent A two parameter distribution was proposed by \cite{mazucheli2018unit} known as unit-Weibull (UW) distribution. This distribution is obtained by making the transformation $X=e^{-Y}$, where random variable Y follows Weibull distribution. Authors shows that this distribution performed better than Kumaraswamy, Beta and other well known distributions. The probability density function (\textit{pdf}) of UW distribution with parameters $\alpha$ and $\beta$  is 
\begin{equation}\label{One_One}
	f(x; \alpha, \beta)=\frac{1}{x}\alpha\beta(-\ln x)^{\beta-1}e^{-\alpha(-\ln x)^{\beta}}, \hspace{0.5cm} 0<x<1\nonumber
\end{equation}
with corresponding cumulative distribution function (cdf) as
\begin{equation}\label{One_Two}
	F(x;\alpha, \beta)=e^{-\alpha(-\ln x)^{\beta}}, \hspace{0.5cm}0<x<1,\nonumber
\end{equation}
where $\alpha$ and $\beta$ are shape parameters.\textit{Pdf} of this transformed model exhibits characteristics such as increasing, decreasing, bathtub-shaped, unimodal, and anti-unimodal. This distribution encompasses a variety of sub-distributions, including instances such as the standard uniform distribution which spans the interval $(0, 1)$ with parameters $\alpha = 1$ and $\beta = 1$, the power function distribution with $\beta = 1$, as well as the unit Rayleigh distribution with $\beta = 2$.\par
For further insights into additional characteristics and properties associated with the UW distribution, interested readers are encouraged to explore the comprehensive study conducted by \cite{mazucheli2018unit}. Following its emergence, different researchers have studied this distribution in diverse scenarios. For instance, \cite{mazucheli2019unit} presented a quantile regression model by considering UW distribution. \cite{iliev2019study} conducted study on the one-side Hausdorff approximation of the Heaviside step function by employing UW, Unit-logistic and Topp Leone cumulative sigmoid functions. \cite{menezes2021bias} presented corrected biased maximum likelihood estimators for the parameters of UW distribution using different methods like Cox-Snell, Parametric bootstrap, and Firth. \cite{alotaibi2020bayesian} delved into the analysis of multicomponent stress strength reliability using both classical and Bayesian approaches, focusing on the UW distribution. \cite{alotaibi2021estimation} obtained estimates of system reliability under known and unknown parameters using classical and Bayesian approaches when the data were observed as progressively type II censored. The authors also constructed asymptotic intervals using the Fisher information matrix, boot-p, and boot-t intervals for system reliability. \cite{almetwally2023analysis} conducted the study on UW distribution for the progressive Type-II censored data. The authors presented estimates for the unknown parameters of UW distribution using two classical methods like MLE and the maximum product of spacing. In the Bayesian estimation paradigm, authors incorporated both the likelihood function and the product of spacing function to estimate the parameters of UW distribution.\par
In this article, we have broadened the study on UW distribution within a dual generalized order statistics (\textit{dgos}) paradigm. The \textit{dgos} was presented as a model containing  sub-model of ordered random variable like, reverse order statistics, lower record values, generalized lower record values, and so on. By taking into account the order statistics of component lifetimes, these models may be used to study the reliability of complex systems, which aids in comprehending how the failure of a single component affects the reliability of the entire system. On the other hand the \textit{dgos} can be utilized in designing redundancy and safety measures in critical system. For example, Consider that it is important to identify and evaluate the most severe failures in decreasing order of severity for a critical system, such as nuclear power plants. This makes it easier to plan safety and redundancy measures appropriately. In this study we provide the practical applicability of UW distribution under the more generalized framework of ordered random variable known as \textit{dgos}. \cite{pawlas2001recurrence} first established \textit{dgos} as a lower generalized order statistic. Following that \cite{burkschat2003dual} presented various distributional properties of \textit{dgos} and their sub-models. Also established the connection between \textit{gos} and \textit{dgos}.Numerous authors have addressed the estimation problem for various distributions within the context of the \textit{dgos} framework. We will delve into some of the prominent research in this area. \cite{jaheen2011bayesian} utilized the \textit{dgos} model for Bayesian estimation of parameters of exponentiated Weibull distribution. They have also utilized the MCMC algorithm to calculate Bayes estimators by considering both symmetric and asymmetric loss functions. \cite{abd2021bayesian} have estimated the parameters of Exponentiated  Generalized Inverted Kumaraswamy Distribution using the Bayesian approach under the \textit{dgos} framework. 
\section{Mathematical Formulation of Problem} 
Let $X_{1}$, $X_{1}$$\dots$,$X_{n}$ be sequence of independent identically distributed random variables with absolutely continuous distribution function (\textit{df}) $F(.)$ and probability density function (\textit{pdf}) $f(.)$. Let $n\in N$, $n\geq2$, $k\geq1$ and $\tilde{m}=\{m_{1}, m_{2},\dots, m_{n-1}\}\in \mathcal{R}^{n-1}$, $M_{r}=\sum_{j=r}^{n-1}m_{j}$, such that $\gamma_{r}=k+n-r+M_{r}>0$ $\forall$ $r\in{1, 2, \dots}, n-1$. Then $X(1, n, \tilde{m}, k)$, $X(2, n, \tilde{m}, k)$,$\dots$,$X(n, n, \tilde{m}, k)$ are called dual generalized order statistics (\textit{dgos}) with joint \textit{pdf}
\begin{equation}\label{Two_One}
	k\left(\prod_{j=1}^{n-1}\gamma_{j}\right)\left(\prod_{i=1}^{n-1}\left[F(x_{i})\right]^{m_{i}}f(x_{i})\right)\left[F(x_{n})\right]^{k-1}f(x_{n})\nonumber
\end{equation}
	for $F^{-1}(1)>x_{1}\geq x_{2}\geq\dots\geq x_{n}>F^{-1}(0)$. The \textit{dgos} is an unified approach that contains several models of ordered random variable arranged in descending order  of magnitude. Taking different values of parameters, $dgos$ can be reduced to different-different models of ordered random variables. For example, when we set $m_{i}=0$ and $k=1$, then the obtained model from $dgos$ is known as reversed order statistics, when set $m_{i}=-1$, then the obtained model from $dgos$ is known as $k^{th}$ or generalized lower record values and when we set $m_{i}=-1$ along with $k=1$, then the obtained model is known as ordinary lower record values, etc. To know more about models of ordered random variables readers are advised to go through the articles/books: \cite{david2004order}, \cite{ahsanullah2004record}, \cite{arnold2008first}.
	Let us suppose that $X_{1}, X_{2},\dots, X_{n}$ be the $n$ \textit{dgos} taking from UW($\alpha$, $\beta$) then the likelihood function written as 
\begin{equation}\label{Two_Two}
	L\left(\alpha, \beta|x\right)=
	\begin{cases}
		k\left(\alpha\beta\right)^{n}\prod_{i=1}^{n}\frac{\left(-\ln x_{i}\right)^{\beta-1}}{x_{i}}\left(\prod_{j=1}^{n-1}\gamma_j\right)e^{-\alpha\left\{\sum_{i=1}^{n-1}\left(m_{i}+1\right)\left(-\ln x_{i}\right)^{\beta}+k\left(-\ln x_{n}\right)^{\beta}\right\}},\\ &\hspace{-0.6in}\alpha>0, \beta>0 \\
		0,  &\hspace{-0.4in}\text{otherwise}
		\end{cases}
\end{equation}
Now we have considered two parameter gamma distributions as independent informative priors for the shape parameters $\alpha$ and $\beta$ 
\begin{equation}\label{prior}
	\left.
	\begin{split}
		&\pi \left(\alpha\right)=\frac{b_{1}^{a_{1}}}{\Gamma a_{1}}\alpha^{a_{1}-1}e^{-b_{1}\alpha}, \hspace{0.3in}a_{1}>0, b_{1}>0, \alpha>0\\
	&\pi \left(\beta\right)=\frac{b_{2}^{a_{2}}}{\Gamma a_{2}}\beta^{a_{2}-1}e^{-b_{2}\beta}, \hspace{0.3in}a_{2}>0, b_{2}>0, \beta>0
	\end{split}
	\right\}.
\end{equation}
The square error loss function (SELF) is defined as 
\begin{equation}
	L_{1}(\hat{\theta}-\theta)=\left(\hat{\theta}-\theta\right)^{2},\hspace{0.3in}\theta>0.\nonumber
\end{equation}
The posterior mean $\hat{\theta}_{SELF}=E\left(\theta|x\right)$ is the Bayes estimator under SELF. The LINEX loss function is defined as
\begin{equation}
	L_2\left(\hat{\theta}, \theta\right)=e^{c\left(\hat{\theta}-\theta\right)}-c\left(\hat{\theta}-\theta\right)-1,\hspace{0.3in}c\neq0 \nonumber
\end{equation}
and Bayes estimator corresponding to LINEX loss function is given as
\begin{equation*}
	\theta_{LINEX}=-\frac{1}{c}\ln\left(E\left(e^{-c\theta}\right)\right).\nonumber
\end{equation*}
The general entropy (GE) function defined as
\begin{equation}
	L_{3}(\hat{\theta}, \theta)\propto \left(\frac{\hat{\theta}}{\theta}\right)^{c}-c\ln\left(\frac{\hat{\theta}}{\theta}\right)-1.\nonumber
\end{equation}
The Bayes estimator corresponding to GE is given as
\begin{equation*}
	\theta_{GE}=\left[E\left(\theta\right)^{-c}\right]^{-\frac{1}{c}}.\nonumber
\end{equation*}
Now, the joint posterior density of $\alpha$ and $\beta$ is obtained by using \eqref{Two_Two} and \eqref{prior}, and is given as
\begin{align}\label{joint_posterior}
\pi\left(\alpha, \beta|x\right)&\propto\alpha^{n+a_{1}-1}\beta^{n+b_{1}-1}\prod_{i=1}^{n}\left(x_{i}\right)^{-1}\left(-\ln x_{i}\right)^{\beta-1} e^{-b_{2}\beta} \nonumber\\
&\qquad \times e^{-\alpha\left\{b_{1}+\sum_{i=1}^{n-1}\left(m_{i}+1\right)\left(-\ln x_{i}\right)^{\beta}+k\left(-\ln x_{n}\right)^{\beta}\right\}} , \qquad\qquad\alpha>0, \beta>0.\nonumber
\end{align}
\section{Bayesian Estimation of the Problem}
For finding the Bayes estimators of $\alpha$, $\beta$ and $R(t)$, we employ Tierney and Kadane’s approximation, Lindley Approximation and Markov chain Monte Carlo methods. The reason for opting these approximation technique is due to the complex form of posterior joint density function.
\subsection{Lindley Approximation } 
Here, we have used Lindley approximation method to approximate the ratio of integrals which was proposed by \cite{lindley1980approximate}. The Bayes estimator for a parameter $\theta$ in presence of any loss function can be represented in the form as given below
\begin{equation}\label{lindley_ratio}
	E\left(\zeta\left(\theta\right)|x\right)=\frac{\int \zeta \left(\theta\right)e^{\mathcal{L\left(\theta\right)}+\phi\left(\theta\right)}d\theta}{\int e^{\mathcal{L\left(\theta\right)}+\phi\left(\theta\right)}},
\end{equation}
where $\mathcal{L}$ is the logarithm of likelihood function and $\phi$ is the logarithm of prior distribution of $\theta$.
In our case $\theta=\left(\alpha, \beta\right)$, thus \eqref{lindley_ratio} can be written as
\begin{equation}
	E\left(\zeta\left(\alpha, \beta \right)|x\right)=\frac{\int \int  \zeta \left(\alpha, \beta\right)e^{\mathcal{L\left(\alpha, \beta\right)}+\phi\left(\alpha, \beta\right)}d\alpha d\beta}{\int \int  e^{\mathcal{L\left(\alpha, \beta\right)}+\phi\left(\alpha, \beta\right)}d\alpha d\beta},\nonumber
\end{equation}
where $\mathcal{L}\left(\alpha, \beta\right)=\ln L\left(\alpha, \beta\right)$ and $\phi\left(\alpha, \beta\right)=\ln\pi\left(\alpha\right)+ \ln\pi\left(\beta\right)$. Thus, using this method, the approximated value of $E\left(\zeta\left(\alpha, \beta\right)|x\right)$ is given as
\begin{equation*}
	E\left(\zeta\left(\alpha, \beta\right)|x\right)\approx\zeta\left(\alpha, \beta \right)+\frac{1}{2}\sum_{i=1}^{2}\sum_{j=1}^{2}\zeta_{ij}\tau_{ij}+\sum_{i=1}^{2}\phi_{i}P_{i}\hspace{2in}
\end{equation*}
\begin{equation}
\hspace{1.5in}	+\frac{1}{2}\sum_{i=1}^{2}\mathcal{L}_{iii}\tau_{ij}P_{i}+\frac{1}{2}\left[\mathcal{L}_{112}\left(2\tau_{12}P_{1}+\tau_{11}P_{2}\right)+\mathcal{L}_{122}\left(\tau_{22}P_{1}+2\tau_{12}P_{2}\right)\right],\nonumber
\end{equation}
where,
\setcounter{equation}{3}
\begin{equation}\label{Three_Four}
	\left.
	\begin{split}
&\zeta_{1}=\frac{\partial\zeta\left(\alpha, \beta\right)}{\partial\alpha},
\zeta_{2}=\frac{\partial\zeta\left(\alpha, \beta\right)}{\partial\beta},
\zeta_{11}=\frac{\partial^{2}\zeta\left(\alpha,\beta\right)}{\partial\alpha^{2}}, \zeta_{22}=\frac{\partial^{2}\zeta\left(\alpha, \beta\right)}{\partial\beta^{2}}, \zeta_{12}=\frac{\partial^{2}\zeta\left(\alpha, \beta\right)}{\partial\alpha\partial\beta}=\zeta_{21}\\
&\mathcal{L}_{11}=\frac{\partial^{2}\mathcal{L}\left(\alpha, \beta|x\right)}{\partial\alpha^{2}}, \mathcal{L}_{22}=\frac{\partial^{2}\mathcal{L}\left(\alpha, \beta|x\right)}{\partial\beta^{2}},
\mathcal{L}_{112}=\frac{\partial^{3}\mathcal{L}\left(\alpha, \beta|x\right)}{\partial\alpha^{2}\partial\beta},
\mathcal{L}_{111}=\frac{\partial^{3}\mathcal{L}\left(\alpha, \beta|x\right)}{\partial\alpha^{3}}\\
&\mathcal{L}_{222}=\frac{\partial^{3}\mathcal{L}\left(\alpha, \beta|x\right)}{\partial\beta^{3}}, \phi_{1}=\frac{\partial\phi\left(\alpha, \beta\right)}{\partial\alpha},
\phi_{2}=\frac{\partial\phi\left(\alpha, \beta\right)}{\partial\beta}, P_{r}=\sum_{j=1}^{2}\zeta_{j}\tau_{rj}
       	\end{split}
	\right\}.
\end{equation}
where $\tau_{rj}$ is the $\left(r,j\right)^{th}$ element of the inverse matrix $\left[-\mathcal{L}_{ij}\right]$. We have obtained the Bayes estimator for the parameters $\alpha$ and $\beta$ using the quantities represented in (\ref{Three_Four}) and by replacing with their respective MLEs.\par
\hspace*{0.2in}For the UW distribution, the terms as given in (\ref{Three_Four}) are reduced in the form as:
\setcounter{equation}{4}
\begin{equation}\label{Three_Five}
	\left.
	\begin{split}
		&\mathcal{L}_{22}=-\frac{n}{\beta^{2}}-\alpha\left[\sum_{i=1}^{n-1}\left(m_{i}+1\right)\left(-\ln x_{i}\right)^{\beta}\left(\ln\left(-\ln x_{i}\right)\right)^{2}+k\left(-\ln x_{n}\right)^{\beta}\left(\ln\left(-\ln x_{n}\right)\right)^{2}\right]\\
		&\mathcal{L}_{12}=-\left[\sum_{i=1}^{n-1}\left(m_{i}+1\right)\left(-\ln x_{i}\right)^{\beta}\ln\left(-\ln x_{i}\right)+k\left(-\ln x_{n}\right)^{\beta}\ln\left(-\ln x_{n}\right)\right]\\
		&\mathcal{L}_{122}=-\left[\sum_{i=1}^{n-1}\left(m_{i}+1\right)\left(-\ln x_{i}\right)^{\beta}\left(\ln\left(-\ln x_{i}\right)\right)^{2}+k\left(-\ln x_{n}\right)^{\beta}\left(\ln\left(-\ln{x_{n}}\right)\right)^{2}\right]\\
		&\mathcal{L}_{222}=\frac{2n}{\beta^{3}}-\alpha\left[\sum_{i=1}^{n-1}\left(m_{i}+1\right)\left(-\ln x_{i}\right)^{\beta}\left(\ln\left(-\ln x_{i}\right)\right)^{3}+k\left(-\ln x_{n}\right)^{\beta}\left(\ln\left(-\ln x_{n}\right)\right)^{3}\right]\\
		&\mathcal{L}_{11}=-\frac{n}{\alpha^{2}}, \mathcal{L}_{112}=0,  \mathcal{L}_{111}=\frac{2n}{\alpha^{3}}, \phi_{1}=\frac{a_{1}-1}{\alpha}-b_{1}, \phi_{2}=\frac{a_{2}-1}{\beta}-b_{2}
\end{split}
\right\}.
\end{equation}
Utilizing quantities given in (\ref{Three_Five}), The Bayes estimators for unknown parameters can be obtained. Here, all quantities except $\zeta\left(\alpha, \beta\right)$ and its derivative, are same for each of the Bayes estimator. For example in the case of SELF, the quantity for Bayes estimator of $\alpha$ can be represented as,
\begin{equation*}
	\zeta\left(\alpha, \beta\right)=\alpha,~\zeta_{1}=1,~ \zeta_{2}=0=\zeta_{11}=\zeta_{12}=\zeta_{22}=\zeta_{21}.
\end{equation*}
For the Bayes estimator $\beta$,
\begin{equation*}
	\zeta\left(\alpha, \beta\right)=\beta,~\zeta_{2}=1,~ \zeta_{1}=0=\zeta_{11}=\zeta_{12}=\zeta_{22}=\zeta_{21}.
\end{equation*}
For the Bayes estimator of $R\left(t\right)$,
\begin{align*}
	&\zeta\left(\alpha, \beta\right)=1-e^{-\alpha\left(-\ln t\right)^{\beta}},~\zeta_{1}=\left(-\ln t\right)^{\beta}e^{-\alpha\left(-\ln t\right)^{\beta}}, \zeta_{2}=\alpha\left(-\ln t\right)^{\beta}\ln\left(-\ln t\right)e^{-\alpha\left(-\ln t\right)^{\beta}}\\
	&\zeta_{11}=\left(-\ln t\right)^{2\beta}e^{-\alpha\left(-\ln t\right)^{\beta}},~\zeta_{22}=\alpha\left(\ln \left(-\ln t\right)\right)^{2}\left(-\ln t\right)^{\beta}e^{-\alpha\left(-\ln t\right)^{\beta}}\left(1-\alpha\left(-\ln t\right)^{\beta}\right)\\
	&\zeta_{12}=\left(-\ln t\right)^{\beta}\ln\left(-\ln t\right)e^{-\alpha\left(-\ln t\right)^{\beta}}\left(1-\alpha\left(-\ln t\right)^{\beta}\right)=\zeta_{21}.
\end{align*}
Similarly following the same idea for LINEX and GE loss function and for $\zeta\left(\alpha, \beta\right)$ given in Table [\ref{estim_table}], the Bayes estimators and their derivatives can easily be found as discussed for SELF.
\begin{table}[h]
	\centering
	\caption{Quantity $\zeta\left(\alpha, \beta\right)$ for different loss functions in Lindley method }
	\label{estim_table}
\begin{tabular}{l|l|l}
	\toprule
	Parameter & LINEX & GE\\ 
	\midrule
	$\alpha$ & $e^{-c\alpha}$ & $\alpha^{-c}$ \\

	$\beta$ & $e^{-c\beta}$ & $\beta^{-c}$ \\

	$R(t)$ & $\exp\left(-c(1-e^{-\alpha(-\ln t)^\beta})\right)$ & $(1-e^{-\alpha(-\ln t)^\beta})^{-c}$\\
	\bottomrule
\end{tabular}
\end{table}
\subsection{Tierney and Kadane's approximation}
The Tierney and Kadane's approximation (T-K) method is an another approach to evaluate the ratio integral, which was introduced by \cite{tierney1986accurate}. Under this method, The posterior expectation $E\left(\zeta\left(\alpha, \beta\right)|x\right)$ can be written as
\begin{equation*}
	E\left[\zeta\left(\alpha, \beta\right)|x\right]=\frac{\int e^{n\psi^{*}\left(\alpha, \beta\right)}d\alpha d\beta}{\int e^{n\psi\left(\alpha, \beta\right)}d\alpha d\beta}.
\end{equation*}
where, $\psi\left(\alpha, \beta\right)=\left(\mathcal{L}\left(\alpha, \beta\right)+\phi\left(\alpha, \beta\right)\right)/n$ and $\psi^{*}\left(\alpha, \beta\right)=\psi\left(\alpha, \beta\right)+\frac{1}{n}\ln\zeta\left(\alpha, \beta\right).$\\
Here, $\mathcal{L}\left(\alpha, \beta\right)$ and $\phi\left(\alpha, \beta\right)$ same as defined in the last section.
Thus, the approximate value of posterior expectation of  $\zeta\left(\alpha, \beta\right)$ using T-K approximation method can be written as
\begin{equation}\label{tk_bayes}
	\hat{E}\left[\zeta\left(\alpha, \beta\right)\right]=\left(\frac{\det{\Sigma^{*}}}{\det{\Sigma}}\right)^{1/2}e^{n\left\{\psi^{*}\left(\hat{\alpha}^{*}, \hat{\beta}^{*}\right)-\psi\left(\hat{\alpha},\hat{\beta}\right)\right\}},
\end{equation}
where, $\left(\hat{\alpha}, \hat{\beta}\right)$ and $\left(\hat{\alpha}^{*}, \hat{\beta}^{*}\right)$ maximizes $\psi\left(\hat{\alpha},\hat{\beta}\right)$ and $\psi^{*}\left(\hat{\alpha}^{*}, \hat{\beta}^{*}\right)$ respectively. $\Sigma$ and $\Sigma^{*}$ are the inverse hessian matrices of $\psi\left(\alpha, \beta\right)$ and $\psi^{*}\left(\alpha^{*}, \beta^{*}\right)$ at $\left(\hat{\alpha}, \hat{\beta}\right)$ and $\left(\hat{\alpha}^{*}, \hat{\beta}^{*}\right)$, respectively.
In our case, we have
\begin{align}
	&\psi\left(\alpha, \beta\right)=\frac{1}{n}\left[C+\left(n+a_{1}-1\right)\ln\alpha+\left(n+a_{2}-1\right)\ln\beta-\alpha\left\{\sum_{i=1}^{n-1}\left(m_{i}+1\right)\left(-\ln x_{i}\right)^{\beta}\right.\right.\nonumber\\
	&\qquad\qquad+k\left.\left.\left(-\ln x_{n}\right)^{\beta}\right\}\right]+\frac{1}{n}\left[\left(\beta-1\right)\sum_{i=1}^{n}\ln\left(-\ln x_{i}\right)-\sum_{i=1}^{n}\ln x_{i}-\alpha b_{1}-\beta b_{2}\right]\nonumber
\end{align}
\begin{align}
	&\psi^{*}\left(\alpha, \beta\right)=\frac{1}{n}\left[C+\left(n+a_{1}-1\right)\ln\alpha+\left(n+a_{2}-1\right)\ln\beta-\alpha\left\{\sum_{i=1}^{n-1}\left(m_{i}+1\right)\left(-\ln x_{i}\right)^{\beta}\right.\right.\nonumber\\
	&\qquad\qquad\left.\left.+k\left(-\ln x_{n}\right)^{\beta}\right\}\right]+\frac{1}{n}\left[\left(\beta-1\right)\sum_{i=1}^{n}\ln\left(-\ln x_{i}\right)-\sum_{i=1}^{n}\ln x_{i}-\alpha b_{1}-\beta b_{2}+\zeta\left(\alpha, \beta\right)\right],\nonumber
\end{align}
where $C$ is a constant independent of parameters $\alpha$, $\beta$. Further, obtain $	\psi_{1}= \frac{\partial}{\partial\alpha}\psi\left(\alpha, \beta\right)$ and $	\psi_{2}= \frac{\partial}{\partial\beta}\psi\left(\alpha, \beta\right),$ then equate both of these quantities to 0. We get
\begin{align}
&\frac{1}{\alpha}-\frac{1}{n}\left[\sum_{i=1}^{n-1}\left(m_{i}+1\right)\left(-\ln x_{i}\right)^{\beta}+k\left(-\ln x_{n}\right)^{\beta}\right]=0 \label{mle1}\\
	&\frac{1}{\beta}-\frac{\alpha}{n}\left[\sum_{i=1}^{n-1}\left(m_{i}+1\right)\left(-\ln x_{i}\right)^{\beta}\ln\left(-\ln x_{i}\right) 
	+k\left(-\ln x_{n}\right)^{\beta}\ln\left(-\ln x_{n}\right)\right]\nonumber\\
	&\qquad\qquad\qquad\qquad+\frac{1}{n}\sum_{i=1}^{n}\ln\left(-\ln x_{i}\right)+\frac{a_{2}-1}{\beta}-b_{2}=0.\label{mle2}
\end{align}

Now $\hat{\alpha}$ and $\hat{\beta}$ can be obtained by solving \eqref{mle1} and \eqref{mle2}. Thus, the determinant for the negative of the inverse Hessian of $\psi\left(\alpha, \beta\right)$ evaluated at $\left(\hat{\alpha}, \hat{\beta}\right)$ is as:
\begin{equation*}
	\det\Sigma=\left(\psi_{11}\psi_{22}-\psi_{12}^{2}\right)^{-1},
\end{equation*}
where
\begin{align*}
\psi_{11}&= \frac{\partial^{2}}{\partial\alpha^{2}}\psi\left(\alpha, \beta\right)\bigg|_{\hat{\alpha},\hat{\beta}}=-\frac{1}{\hat{\alpha}^{2}}-\frac{a_{1}-1}{\hat{\alpha}^{2}}\\
\psi_{12}&= \frac{\partial^{2}}{\partial\alpha\partial\beta}\psi\left(\alpha, \beta\right)\bigg\vert_{\hat{\alpha}, \hat{\beta}}=\psi_{21} =-\frac{1}{n}\left[\sum_{i=1}^{n}\left(m_{i}+1\right)\left(-\ln x_{i}\right)^{\hat{\beta}}\ln\left(-\ln x_{i}\right) \right.\nonumber\\
&\qquad\qquad \qquad\qquad \qquad\qquad\qquad\qquad\left.+k\left(-\ln x_{n}\right)^{\hat{\beta}}\ln\left(-\ln x_{n}\right)\right]\\
\psi_{22}&= \frac{\partial^{2}}{\partial{\beta^{2}}}\psi\left(\alpha, \beta\right)\bigg|_{\hat{\alpha}, \hat{\beta}}=-\frac{1}{\hat{\beta^{2}}}-\frac{\hat{\alpha}}{n}\left[\sum_{i=1}^{n-1}\left(m_{i}+1\right)\left(-\ln x_{i}\right)^{\hat{\beta}}\left(\ln\left(-\ln x_{i}\right)^{2}\right)\right. \nonumber\\
&\qquad\qquad \qquad\qquad \qquad\qquad\qquad\qquad\left. + k\left(-\ln x_{n}\right)^{\hat{\beta}}\left(\ln\left(-\ln x_{n}\right)\right)\right]-\frac{a_{2}-1}{\hat{\beta^{2}}}.
\end{align*}
The determinant of $\Sigma$, i.e., $\left(\psi_{11}\psi_{22}-\psi_{12}^{2}\right)^{-1}$  will be same for estimates of parameters under all loss functions while determinant of $\Sigma^{*}$, i.e., $\left(\psi_{11}^{*}\psi_{22}^{*}-\psi_{12}^{*2}\right)^{-1}$ will not be same as it contains the term $\frac{1}{n}\ln\zeta\left(\alpha, \beta\right)$ which is different for each parameter and loss function. \par 
First we will find the Bayes estimators for $\alpha$, $\beta$ and $R(t)$ under SELF. For Bayes estimator of $\alpha$, we have $\zeta(\alpha,\beta)=\alpha$ this implies

\begin{equation}\label{tk_baye_alpha_self}
		\psi^{*}\left(\alpha, \beta\right)=\psi\left(\alpha, \beta\right)+\frac{1}{n}\ln\alpha.
\end{equation}
Now, differentiating \eqref{tk_baye_alpha_self} with respect to $\alpha$, $\beta$, we get
\begin{align*}
	\frac{\partial\psi\left(\alpha, \beta\right)}{\partial\alpha}+\frac{1}{n\alpha}=0\qquad \& \qquad\frac{\partial\psi\left(\alpha, \beta\right)}{\partial\beta}=0.
\end{align*}
Thus, the solution of first derivatives given above produces  the MLE's for $\alpha$ and $\beta$. Now, determinant for the negative of the inverse hessian matrix
of $\psi^{*}\left(\alpha, \beta\right)$ obtained at $\hat{\alpha}^*$ and $\hat{\beta}^*$ is  $\det\Sigma^{*}=\left(\psi^{*}_{11}\psi^{*}_{22}-\psi^{*2}_{12}\right)^{-1}$. The quantities $\psi^{*}_{11},$ $\psi^{*}_{12},$ and $\psi^{*}_{22}$ can be obtained by taking derivatives of $\psi^{*}\left(\alpha, \beta\right)$ as in the case of $\det \Sigma.$ Using the quantity $\det \Sigma$ and $\det \Sigma^*$ in equation \eqref{tk_bayes}, we get the Bayes estimate of $\alpha.$\par 
For the Bayes estimator of $\beta$, we have $\zeta\left(\alpha, \beta\right)=\beta$, this implies

\begin{equation}\label{tk_baye_beta_self}
	\psi^{*}\left(\alpha, \beta\right)=\psi\left(\alpha, \beta\right)+\frac{1}{n}\ln\beta.
\end{equation}
Now, differentiating \eqref{tk_baye_beta_self} with respect to $\alpha$, $\beta$, we get
\begin{align*}
	\frac{\partial\psi\left(\alpha, \beta\right)}{\partial\alpha}=0\qquad \& \qquad\frac{\partial\psi\left(\alpha, \beta\right)}{\partial\beta}+\frac{1}{n\beta}=0.
\end{align*}
Thus, the solution of first derivatives given above produces  the MLE's for $\alpha$ and $\beta$. Now, determinant for the negative of the inverse hessian matrix
of $\psi^{*}\left(\alpha, \beta\right)$ obtained at $\hat{\alpha}^*$ and $\hat{\beta}^*$ is  $\det\Sigma^{*}=\left(\psi^{*}_{11}\psi^{*}_{22}-\psi^{*2}_{12}\right)^{-1}$. Using the same idea of Bayes estimator of $\alpha$, we find the Bayes estimator of $\beta$. \par
Following in the same steps, we can find the Bayes estimator of $R(t)$. For the Bayes estimator of $R\left(t\right)$, we have $\zeta\left(\alpha, \beta\right)=1-e^{-\alpha\left(-\ln t\right)^{\beta}}$, this implies
\begin{align}\label{tk_baye_rel_self}
	\psi^{*}\left(\alpha, \beta\right)=\psi\left(\alpha, \beta\right)+\frac{1}{n}\ln\left(1-e^{-\alpha\left(-\ln t\right)^{\beta}}\right).
\end{align}
Now, differentiating \eqref{tk_baye_rel_self} with respect to $\alpha$, $\beta$, we get
\begin{align*}
	&\frac{\psi\left(\alpha, \beta\right)}{\partial\alpha}+\frac{\left(-\ln t\right)^{\beta}e^{\left(-\ln t\right)^{\beta}}}{n\left(1-e^{-\alpha\left(-\ln t\right)^{\beta}}\right)}=0\\
	&\frac{\psi\left(\alpha, \beta\right)}{\partial\beta}+\frac{\alpha\left(-\ln t\right)^{\beta}\ln\left(-\ln t\right)e^{\left(-\ln t\right)^{\beta}}}{n\left(1-e^{-\alpha\left(-\ln t\right)^{\beta}}\right)}=0.
\end{align*}
Thus, the solution of first derivatives given above produces  the MLE's for $\alpha$ and $\beta$. Now, determinant for the negative of the inverse hessian matrix
of $\psi^{*}\left(\alpha, \beta\right)$ obtained at $\hat{\alpha}^*$ and $\hat{\beta}^*$ is  $\det\Sigma^{*}=\left(\psi^{*}_{11}\psi^{*}_{22}-\psi^{*2}_{12}\right)^{-1}$, where

\begin{align*}
	&\psi^{*}_{11}=\frac{\partial^{2}\psi^{*}\left(\alpha, \beta\right)}{\partial\alpha^{2}}=\psi_{11}-\frac{\left(-\ln t\right)^{2\hat{\beta}}e^{-\alpha\left(-\ln t\right)^{\hat{\beta}}}}{n\left(1-e^{-\hat{\alpha}\left(-\ln t\right)^{\hat{\beta}}}\right)^{2}}\\
	&\psi^{*}_{12}=\frac{\partial\psi^{*}\left(\alpha, \beta\right)}{\partial\alpha\partial\beta}=\psi_{12}+\frac{\left(-\ln t\right)^{\hat{\beta}}\ln \left(-\ln t\right)\left[\left(\left(\hat{\alpha}\left(-\ln t\right)^{\hat{\beta}}\right)-1\right)e^{\hat{\alpha}\left(-\ln t\right)^{\hat{\beta}}}+1\right]}{n\left(1-e^{-\hat{\alpha}\left(-\ln t\right)^{\hat{\beta}}}\right)^{2}}\\
	&\psi^{*}_{22}=\frac{\partial\psi^{*}\left(\alpha, \beta\right)}{\partial\beta^{2}}=\psi_{22}+\frac{\hat{\alpha}\left(-\ln t\right)^{\hat{\beta}}\left(\ln \left(-\ln t\right)\right)^{2}\left[\left(\left(\hat{\alpha}\left(-\ln t\right)^{\hat{\beta}}\right)-1\right)e^{\hat{\alpha}\left(-\ln t\right)^{\hat{\beta}}}+1\right]}{n\left(1-e^{-\hat{\alpha}\left(-\ln t\right)^{\hat{\beta}}}\right)^{2}}.
\end{align*}
The quantities given in Table [\ref{tab:tk_estim_table}] are used for finding the Bayes estimators under LINEX and GE loss functions. The steps involved for obtaining the Bayes estimators are similar to SELF.
\begin{table}[h]
	\centering
	\caption{Quantity $\psi^{*}\left(\alpha, \beta\right)$ for different loss functions in T-K method }
	\label{tab:tk_estim_table}
	\begin{tabular}{l|l|l}
		\toprule
		Parameter & LINEX & GE\\ \midrule
			$\alpha$ & $\psi\left(\alpha, \beta\right)-\dfrac{c\alpha}{n}$ & $\psi\left(\alpha, \beta\right)-\dfrac{c\ln\alpha}{n}$ \\
		
		\multirow{2}{*}{$\beta$} & \multirow{2}{*}{$\psi\left(\alpha, \beta\right)-\dfrac{c\beta}{n}$} & \multirow{2}{*}{$\psi\left(\alpha, \beta\right)-\dfrac{c\ln\beta}{n}$} \\
		&&\\
		
			\multirow{2}{*}{$R(t)$} & 	\multirow{2}{*}{$\psi\left(\alpha, \beta\right)-\dfrac{c(1-e^{-\alpha(-\ln t)^\beta})}{n}$} &	\multirow{2}{*}{ $\psi\left(\alpha, \beta\right)-\dfrac{c\ln(1-e^{-\alpha(-\ln t)^\beta})}{n}$}\\
				&&\\
				\bottomrule
	\end{tabular}
\end{table}

\subsection{Markov Chain Monte Carlo}
Now, we use Markov Chain Monte Carlo (MCMC) method to obtain the Bayes estimators of the unknown parameters of  UW distribution. The idea behind this to approximate the posterior distribution of parameters in the context of Bayesian analysis. The MCMC method is used to generate random sample from the posterior distribution and then use the generated data to obtain the Bayes estimator across the considered loss functions. For this, first we calculate the marginal densities of unknown parameters using the likelihood function and prior distribution. The marginal densities of $\alpha$ and $\beta$ as
\begin{align}\label{post.alpha}
	\left. \begin{array}{cl}
		\pi(\alpha|\beta,\boldsymbol{x})\propto&\textup{Gamma}\left(n+a_1,b_1+ \sum_{i=1}^{n-1}\left(m_{i}+1\right)\left(-\ln x_{i}\right)^{\beta}+k\left(-\ln x_{n}\right)^{\beta}\right)\\
		\pi(\beta|\alpha,\boldsymbol{x})\propto& \beta^{n+a_2-1}\prod_{i=1}^n\left(\ln x_i^{\beta}\right) e^{-b_2\beta-\alpha\left\{\sum_{i=1}^{n-1}\left(m_{i}+1\right)\left(-\ln x_{i}\right)^{\beta}+k\left(-\ln x_{n}\right)^{\beta}\right\}}
	\end{array}\right\rbrace.
\end{align}
 From \eqref{post.alpha} we observe that the generation of random sample for $\alpha$ is easy as it has a nice closed form of gamma distribution (\cite{geman1987stochastic}). But for $\beta$ it is not easy as it does not have a nice analytical form of any known probability distribution. So for this, we employ the idea of Metropolis Hasting (MH) algorithm with normal distribution (see \cite{gelman2013bayesian}) as the proposal density. The readers are referred to \cite{arshad2023bayesian} to see the algorithm. Once we generate the random data of size $N$ say $\delta^1, \delta^1, \ldots, \delta^N$, we can  obtain the Bayes estimators in the following manner. For the squared error loss function $	\delta_{SELF}=\dfrac{1}{T}\sum_{i=1}^{T}\delta^i$, LINEX loss function $	\delta_{LINEX}=-\dfrac{1}{c} \ln \left(\dfrac{1}{T}\sum_{i=1}^{T}e^{-c\delta^i}\right)$ and GE loss function $	\delta_{GE}=\left(\dfrac{1}{T}\sum_{i=1}^{T}(\delta^i)^{-c}\right)^{-1/c}.$
	\section {Simulation Study}
	This section comprises of studying the behavior of the derived estimators on the simulated model. Various configurations of the parameters, sample sizes and priors have been used and reported in this section. Since \textit{dgos} is an umbrella term containing several ordered random variable based models so we have obtained the results for two sub models that is, order statistic	and lower record values. We have seen that for $m_{i}=-1$ and $k=1,$ the \textit{dgos} model reduces to lower record value and for $m_{i}=0$ and $k=1,$ the \textit{dgos} model reduces to order statistics. The performance of Bayes estimators (order statistics and lower record values) are measured using the criteria of risk function. In order to calculate the risk, first we need to generate the random sample of \textit{dgos}. For this purpose the algorithm discussed by \cite{arshad2023bayesian} is considered here.
	Now, after generating the random sample for the both the submodels, we calculate  Bayes estimators for Lindley approximation, T-K method and MCMC techniques and then the respective risks for each estimator are obtained for 1000 replications. The risks of estimators are calculated for the different configurations of parameters. Two configurations  of priors are considered for calculation of Bayes estimators i.e., Prior I $=(a_i,b_i)=(2,2),i=1,2$ and Prior II $=(a_i,b_i)=(0.05,0.05),i=1,2.$ The calculation is performed using R software(\cite{rcore}). In addition to this, the convergence behaviour of generated Markov chain is tested with the aid of  Gelman Rubin (GR) diagnostic (See \cite{gelman2013bayesian}). With GR diagnostic we find that as we increase the number of iterations, the value of shrink reduction factor is getting close to 1. Hence, we conclude that convergence is achieved. The risks of various estimators are reported in Table [\ref{tab:lind.rec}-\ref{tab:mcmc.order}]. From these tables, the following observations are made.
	\begin{itemize}
		\item[(i)] The Table [\ref{tab:lind.rec}] reports risks of Bayes estimates obtained using Lindley Approximation method for lower record values. From the table, it is observed that risks based on asymmetric loss functions (LINEX and GELF) are much smaller than symmetric loss function. 
		\item[(ii)] The Table [\ref{tab:tk.rec}] reports risks of Bayes estimates obtained using T-K method for lower record values. From the table, it is observed that risks based on asymmetric loss functions (LINEX and GELF) are much smaller than symmetric loss function. It is also observed that risks of estimators based on T-K method are smaller than risk of estimators based on Lindley method.  
		\item[(iii)] The Table [\ref{tab:mcmc.rec}] reports risks of Bayes estimates obtained using MCMC method for lower record values. From the table, it is observed that risks based on asymmetric loss functions (LINEX and GELF) are much smaller than symmetric loss function. It is also observed that mostly risks of estimators based on MCMC method are not smaller than risk of estimators based on T-K method. 
		\item[(iv)] The Table [\ref{tab:lind.order}-\ref{tab:mcmc.order}] report risks of Bayes estimates obtained using Lindley, T-K and MCMC method for order statistics, respectively. Similar observations are seen for risks of all estimators for order statistics as these were for lower record values.
		\item[(v)] From all the tables, it is observed that the risks of all estimators are decreasing as we increase the sample size irrespective of ordered random models. Also, Prior I seems to have shown lesser risk that Prior II.
		\item[(vi)] From these observations it is evident that Bayes estimators based on asymmetric loss functions (LINEX and GELF) are performing better based on their risks. So, In practical scenarios where the underlying assumptions considered in this study are satisfied, it is recommended to use asymmetric loss functions as it provides more flexibility to the model. Also, estimators based on T-K and MCMC methods are performing better than Lindley estimators. 
	\end{itemize} 
	
	\begin{table}[htbp]
		\centering
		\caption{Risks of Lindley Bayes estimates of unknown quantities based on record values for $t= 0.5$}
		\begin{tabular}{cc|cccc|ccc|ccc}
			\toprule
			\multicolumn{2}{c|}{\multirow{2}[4]{*}{$(\alpha,\beta)$}} & \multicolumn{1}{c|}{\multirow{2}[4]{*}{$n$}} & \multicolumn{3}{c|}{SELF} & \multicolumn{3}{c|}{LINEX} & \multicolumn{3}{c}{GELF } \\
			\cmidrule{4-12}    \multicolumn{2}{c|}{} & \multicolumn{1}{c|}{} & $\hat{\alpha}_{risk}$ & $\hat{\beta}_{risk}$ & $\hat{R(t)}_{risk}$ & $\hat{\alpha}_{risk}$ & $\hat{\beta}_{risk}$ & $\hat{R(t)}_{risk}$ & $\hat{\alpha}_{risk}$ & $\hat{\beta}_{risk}$ & \multicolumn{1}{c|}{$\hat{R(t)}_{risk}$} \\
			\midrule
			\multicolumn{12}{c}{$(a_1,a_2,b_1,b_2)=(2,2,2,2)$} \\
			\midrule
			\multicolumn{2}{c|}{\multirow{3}[2]{*}{(1,1)}} & \multicolumn{1}{c|}{5} & 0.2420 & 0.1380 & 0.0500 & 0.0343 & 0.0207 & 0.0087 & 0.0378 & 0.0125 & 0.0101 \\
			\multicolumn{2}{c|}{} & \multicolumn{1}{c|}{10} & 0.1123 & 0.0804 & 0.0171 & 0.0129 & 0.0074 & 0.0024 & 0.0203 & 0.0030 & 0.0053 \\
			\multicolumn{2}{c|}{} & \multicolumn{1}{c|}{15} & 0.0717 & 0.0377 & 0.0086 & 0.0053 & 0.0019 & 0.0011 & 0.0114 & 0.0011 & 0.0031 \\
			\midrule
			\multicolumn{2}{c|}{\multirow{3}[2]{*}{(1.5,1)}} & \multicolumn{1}{c|}{5} & 0.4072 & 0.1651 & 0.0468 & 0.0360 & 0.0210 & 0.0065 & 0.0616 & 0.0116 & 0.0123 \\
			\multicolumn{2}{c|}{} & \multicolumn{1}{c|}{10} & 0.3535 & 0.0888 & 0.0301 & 0.0261 & 0.0085 & 0.0035 & 0.0414 & 0.0066 & 0.0072 \\
			\multicolumn{2}{c|}{} & \multicolumn{1}{c|}{15} & 0.3027 & 0.0416 & 0.0234 & 0.0230 & 0.0044 & 0.0028 & 0.0300 & 0.0044 & 0.0046 \\
			\midrule
			\multicolumn{2}{c|}{\multirow{3}[2]{*}{(1,1.5)}} & \multicolumn{1}{c|}{5} & 0.3699 & 0.3627 & 0.0830 & 0.0333 & 0.0596 & 0.0122 & 0.0437 & 0.0387 & 0.0097 \\
			\multicolumn{2}{c|}{} & \multicolumn{1}{c|}{10} & 0.2016 & 0.2756 & 0.0285 & 0.0351 & 0.0179 & 0.0045 & 0.0261 & 0.0109 & 0.0058 \\
			\multicolumn{2}{c|}{} & \multicolumn{1}{c|}{15} & 0.1219 & 0.1457 & 0.0145 & 0.0142 & 0.0059 & 0.0019 & 0.0166 & 0.0041 & 0.0040 \\
			\midrule
			\multicolumn{2}{c|}{\multirow{3}[2]{*}{(1.5,1.5)}} & \multicolumn{1}{c|}{5} & 0.3586 & 0.2523 & 0.0524 & 0.0278 & 0.0387 & 0.0065 & 0.0665 & 0.0249 & 0.0122 \\
			\multicolumn{2}{c|}{} & \multicolumn{1}{c|}{10} & 0.2338 & 0.1606 & 0.0190 & 0.0177 & 0.0089 & 0.0025 & 0.0337 & 0.0057 & 0.0061 \\
			\multicolumn{2}{c|}{} & \multicolumn{1}{c|}{15} & 0.1848 & 0.0749 & 0.0118 & 0.0110 & 0.0038 & 0.0014 & 0.0200 & 0.0034 & 0.0036 \\
			\midrule
			\multicolumn{12}{c}{$(a_1,a_2,b_1,b_2)=(0.05,0.05,0.05,0.05)$} \\
			\midrule
			\multicolumn{2}{c|}{\multirow{3}[2]{*}{(1,1)}} & 5  & 0.3207 & 0.1897 & 0.0572 & 0.0340 & 0.0243 & 0.0069 & 0.0500 & 0.0226 & 0.0126 \\
			\multicolumn{2}{c|}{} & 10 & 0.2641 & 0.1302 & 0.0451 & 0.0296 & 0.0163 & 0.0055 & 0.0419 & 0.0149 & 0.0098 \\
			\multicolumn{2}{c|}{} & 15 & 0.2398 & 0.0838 & 0.0369 & 0.0275 & 0.0103 & 0.0045 & 0.0350 & 0.0095 & 0.0076 \\
			\midrule
			\multicolumn{2}{c|}{\multirow{3}[2]{*}{(1.5,1)}} & 5  & 0.5144 & 0.1920 & 0.0718 & 0.0537 & 0.0241 & 0.0089 & 0.0881 & 0.0211 & 0.0164 \\
			\multicolumn{2}{c|}{} & 10 & 0.4083 & 0.1354 & 0.0529 & 0.0438 & 0.0171 & 0.0067 & 0.0710 & 0.0158 & 0.0121 \\
			\multicolumn{2}{c|}{} & 15 & 0.3602 & 0.0973 & 0.0460 & 0.0394 & 0.0117 & 0.0058 & 0.0613 & 0.0110 & 0.0099 \\
			\midrule
			\multicolumn{2}{c|}{\multirow{3}[2]{*}{(1,1.5)}} & 5  & 0.2984 & 0.1138 & 0.0488 & 0.0322 & 0.0142 & 0.0058 & 0.0496 & 0.0169 & 0.0108 \\
			\multicolumn{2}{c|}{} & 10 & 0.2634 & 0.0842 & 0.0398 & 0.0302 & 0.0104 & 0.0047 & 0.0415 & 0.0113 & 0.0082 \\
			\multicolumn{2}{c|}{} & 15 & 0.2532 & 0.0693 & 0.0370 & 0.0289 & 0.0087 & 0.0044 & 0.0367 & 0.0090 & 0.0070 \\
			\midrule
			\multicolumn{2}{c|}{\multirow{3}[2]{*}{(1.5,1.5)}} & 5  & 0.5417 & 0.1179 & 0.0683 & 0.0559 & 0.0142 & 0.0083 & 0.0907 & 0.0169 & 0.0155 \\
			\multicolumn{2}{c|}{} & 10 & 0.4180 & 0.0855 & 0.0513 & 0.0449 & 0.0107 & 0.0064 & 0.0728 & 0.0115 & 0.0116 \\
			\multicolumn{2}{c|}{} & 15 & 0.3510 & 0.0705 & 0.0413 & 0.0381 & 0.0089 & 0.0052 & 0.0596 & 0.0094 & 0.0089 \\
			\bottomrule
		\end{tabular}%
		\label{tab:lind.rec}%
	\end{table}%
	\begin{table}[htbp]
		\centering
		\caption{Risks of T-K Bayes estimates of unknown quantities based on record values ($t= 0.5$)}
		\begin{tabular}{cc|c|ccc|ccc|ccc}
			\toprule
			\multicolumn{2}{c|}{\multirow{2}[4]{*}{$(\alpha,\beta)$}} & \multirow{2}[4]{*}{$n$} & \multicolumn{3}{c|}{SELF} & \multicolumn{3}{c|}{LINEX} & \multicolumn{3}{c}{GELF } \\
			\cmidrule{4-12}    \multicolumn{2}{c|}{} &    & $\hat{\alpha}_{risk}$ & $\hat{\beta}_{risk}$ & $\hat{R(t)}_{risk}$ & $\hat{\alpha}_{risk}$ & $\hat{\beta}_{risk}$ & $\hat{R(t)}_{risk}$ & $\hat{\alpha}_{risk}$ & $\hat{\beta}_{risk}$ & \multicolumn{1}{c}{$\hat{R(t)}_{risk}$} \\
			\midrule
			\multicolumn{12}{c}{$(a_1,a_2,b_1,b_2)=(2,2,2,2)$} \\
			\midrule
			\multicolumn{2}{c|}{\multirow{3}[2]{*}{(1,1)}} & 5  & 0.1054 & 0.0937 & 0.0117 & 0.0122 & 0.0105 & 0.0014 & 0.0569 & 0.0000 & 0.0177 \\
			\multicolumn{2}{c|}{} & 10 & 0.0992 & 0.0397 & 0.0120 & 0.0116 & 0.0046 & 0.0015 & 0.0291 & 0.0000 & 0.0161 \\
			\multicolumn{2}{c|}{} & 15 & 0.1127 & 0.0260 & 0.0130 & 0.0131 & 0.0031 & 0.0017 & 0.0323 & 0.0001 & 0.0161 \\
			\midrule
			\multicolumn{2}{c|}{\multirow{3}[2]{*}{(1.5,1)}} & 5  & 0.2304 & 0.1645 & 0.0315 & 0.0292 & 0.0188 & 0.0037 & 0.1231 & 0.0000 & 0.0273 \\
			\multicolumn{2}{c|}{} & 10 & 0.2245 & 0.0721 & 0.0300 & 0.0297 & 0.0085 & 0.0036 & 0.0299 & 0.0034 & 0.0251 \\
			\multicolumn{2}{c|}{} & 15 & 0.2122 & 0.0472 & 0.0272 & 0.0277 & 0.0057 & 0.0033 & 0.0434 & 0.0067 & 0.0218 \\
			\midrule
			\multicolumn{2}{c|}{\multirow{3}[2]{*}{(1,1.5)}} & 5  & 0.1194 & 0.1189 & 0.0089 & 0.0126 & 0.0130 & 0.0011 & 0.0001 & 0.0044 & 0.0115 \\
			\multicolumn{2}{c|}{} & 10 & 0.1064 & 0.0551 & 0.0093 & 0.0114 & 0.0063 & 0.0012 & 0.0036 & 0.0004 & 0.0124 \\
			\multicolumn{2}{c|}{} & 15 & 0.1152 & 0.0431 & 0.0108 & 0.0125 & 0.0051 & 0.0014 & 0.0025 & 0.0062 & 0.0136 \\
			\midrule
			\multicolumn{2}{c|}{\multirow{3}[2]{*}{(1.5,1.5)}} & 5  & 0.1611 & 0.1671 & 0.0181 & 0.0204 & 0.0173 & 0.0021 & 0.0000 & 0.0116 & 0.0210 \\
			\multicolumn{2}{c|}{} & 10 & 0.1647 & 0.1124 & 0.0185 & 0.0214 & 0.0129 & 0.0022 & 0.0420 & 0.0096 & 0.0201 \\
			\multicolumn{2}{c|}{} & 15 & 0.1772 & 0.0692 & 0.0185 & 0.0227 & 0.0080 & 0.0023 & 0.0067 & 0.0023 & 0.0188 \\
			\midrule
			\multicolumn{12}{c}{$(a_1,a_2,b_1,b_2)=(1,1,1,1)$} \\
			\midrule
			\multicolumn{2}{c|}{\multirow{3}[2]{*}{(1,1)}} & 5  & 0.2199 & 0.1413 & 0.0188 & 0.0277 & 0.0149 & 0.0030 & 0.0041 & 0.0247 & 0.0523 \\
			\multicolumn{2}{c|}{} & 10 & 0.1894 & 0.0601 & 0.0190 & 0.0236 & 0.0062 & 0.0030 & 0.0044 & 0.0011 & 0.0321 \\
			\multicolumn{2}{c|}{} & 15 & 0.2063 & 0.0416 & 0.0209 & 0.0246 & 0.0046 & 0.0031 & 0.0122 & 0.0003 & 0.0295 \\
			\midrule
			\multicolumn{2}{c|}{\multirow{3}[2]{*}{(1.5,1)}} & 5  & 0.3543 & 0.2689 & 0.0400 & 0.0440 & 0.0302 & 0.0050 & 0.0607 & 0.0057 & 0.0623 \\
			\multicolumn{2}{c|}{} & 10 & 0.3198 & 0.1056 & 0.0350 & 0.0416 & 0.0121 & 0.0045 & 0.0013 & 0.0081 & 0.0423 \\
			\multicolumn{2}{c|}{} & 15 & 0.3022 & 0.0609 & 0.0300 & 0.0371 & 0.0073 & 0.0038 & 0.0004 & 0.0004 & 0.0274 \\
			\midrule
			\multicolumn{2}{c|}{\multirow{3}[2]{*}{(1,1.5)}} & 5  & 0.2283 & 0.2136 & 0.0159 & 0.0261 & 0.0223 & 0.0024 & 0.0249 & 0.0043 & 0.0421 \\
			\multicolumn{2}{c|}{} & 10 & 0.1949 & 0.1042 & 0.0165 & 0.0224 & 0.0110 & 0.0026 & 0.0527 & 0.0016 & 0.0249 \\
			\multicolumn{2}{c|}{} & 15 & 0.2344 & 0.0770 & 0.0191 & 0.0260 & 0.0085 & 0.0028 & 0.0193 & 0.0004 & 0.0288 \\
			\midrule
			\multicolumn{2}{c|}{\multirow{3}[2]{*}{(1.5,1.5)}} & 5  & 0.3220 & 0.3650 & 0.0285 & 0.0379 & 0.0375 & 0.0034 & 0.0738 & 0.0060 & 0.0523 \\
			\multicolumn{2}{c|}{} & 10 & 0.3013 & 0.1695 & 0.0272 & 0.0375 & 0.0186 & 0.0035 & 0.0274 & 0.0182 & 0.0336 \\
			\multicolumn{2}{c|}{} & 15 & 0.3002 & 0.1018 & 0.0248 & 0.0356 & 0.0116 & 0.0032 & 0.2057 & 0.0260 & 0.0297 \\
			\bottomrule
		\end{tabular}%
		\label{tab:tk.rec}%
	\end{table}%
	\begin{table}[htbp]
		\centering
		\caption{Risks of MCMC Bayes estimates of unknown quantities based on record values ($t= 0.5$)}
		\begin{tabular}{cc|cccc|ccc|ccc}
			\toprule
			\multicolumn{2}{c|}{\multirow{2}[4]{*}{$(\alpha,\beta)$}} & \multicolumn{1}{c|}{\multirow{2}[4]{*}{$n$}} & \multicolumn{3}{c|}{SELF} & \multicolumn{3}{c|}{LINEX} & \multicolumn{3}{c}{GELF } \\
			\cmidrule{4-12}    \multicolumn{2}{c|}{} & \multicolumn{1}{c|}{} & $\hat{\alpha}_{risk}$ & $\hat{\beta}_{risk}$ & $\hat{R(t)}_{risk}$ & $\hat{\alpha}_{risk}$ & $\hat{\beta}_{risk}$ & $\hat{R(t)}_{risk}$ & $\hat{\alpha}_{risk}$ & $\hat{\beta}_{risk}$ & \multicolumn{1}{c}{$\hat{R(t)}_{risk}$} \\
			\midrule
			\multicolumn{12}{c}{$(a_1,a_2,b_1,b_2)=(2,2,2,2)$} \\
			\midrule
			\multicolumn{2}{c|}{\multirow{3}[2]{*}{(1,1)}} & \multicolumn{1}{c|}{5} & 0.1460 & 0.1129 & 0.0159 & 0.0166 & 0.0113 & 0.0023 & 0.0285 & 0.0079 & 0.0255 \\
			\multicolumn{2}{c|}{} & \multicolumn{1}{c|}{10} & 0.1386 & 0.0468 & 0.0162 & 0.0153 & 0.0054 & 0.0020 & 0.0268 & 0.0044 & 0.0219 \\
			\multicolumn{2}{c|}{} & \multicolumn{1}{c|}{15} & 0.1193 & 0.0272 & 0.0158 & 0.0150 & 0.0036 & 0.0021 & 0.0234 & 0.0028 & 0.0193 \\
			\midrule
			\multicolumn{2}{c|}{\multirow{3}[2]{*}{(1.5,1)}} & \multicolumn{1}{c|}{5} & 0.2543 & 0.1895 & 0.0327 & 0.0342 & 0.0189 & 0.0043 & 0.0368 & 0.0109 & 0.0297 \\
			\multicolumn{2}{c|}{} & \multicolumn{1}{c|}{10} & 0.2507 & 0.0810 & 0.0327 & 0.0326 & 0.0092 & 0.0043 & 0.0365 & 0.0061 & 0.0272 \\
			\multicolumn{2}{c|}{} & \multicolumn{1}{c|}{15} & 0.2452 & 0.0480 & 0.0303 & 0.0319 & 0.0074 & 0.0043 & 0.0363 & 0.0051 & 0.0262 \\
			\midrule
			\multicolumn{2}{c|}{\multirow{3}[2]{*}{(1,1.5)}} & \multicolumn{1}{c|}{5} & 0.1878 & 0.1280 & 0.0159 & 0.0180 & 0.0138 & 0.0018 & 0.0181 & 0.0076 & 0.0161 \\
			\multicolumn{2}{c|}{} & \multicolumn{1}{c|}{10} & 0.1629 & 0.0772 & 0.0144 & 0.0176 & 0.0080 & 0.0017 & 0.0177 & 0.0036 & 0.0157 \\
			\multicolumn{2}{c|}{} & \multicolumn{1}{c|}{15} & 0.1346 & 0.0475 & 0.0124 & 0.0169 & 0.0058 & 0.0018 & 0.0171 & 0.0027 & 0.0155 \\
			\midrule
			\multicolumn{2}{c|}{\multirow{3}[2]{*}{(1.5,1.5)}} & \multicolumn{1}{c|}{5} & 0.2183 & 0.1804 & 0.0222 & 0.0266 & 0.0175 & 0.0029 & 0.0259 & 0.0079 & 0.0223 \\
			\multicolumn{2}{c|}{} & \multicolumn{1}{c|}{10} & 0.2041 & 0.1096 & 0.0208 & 0.0250 & 0.0131 & 0.0027 & 0.0231 & 0.0044 & 0.0208 \\
			\multicolumn{2}{c|}{} & \multicolumn{1}{c|}{15} & 0.1997 & 0.0787 & 0.0199 & 0.0246 & 0.0099 & 0.0025 & 0.0218 & 0.0033 & 0.0189 \\
			\midrule
			\multicolumn{12}{c}{$(a_1,a_2,b_1,b_2)=(0.05,0.05,0.05,0.05)$} \\
			\midrule
			\multicolumn{2}{c|}{\multirow{3}[2]{*}{(1,1)}} & 5  & 0.5205 & 0.4258 & 0.0508 & 0.0584 & 0.0497 & 0.0064 & 0.1505 & 0.0236 & 0.1609 \\
			\multicolumn{2}{c|}{} & 10 & 0.5178 & 0.1950 & 0.0481 & 0.0424 & 0.0228 & 0.0058 & 0.1290 & 0.0125 & 0.1296 \\
			\multicolumn{2}{c|}{} & 15 & 0.4821 & 0.0969 & 0.0438 & 0.0435 & 0.0115 & 0.0051 & 0.0920 & 0.0079 & 0.0842 \\
			\midrule
			\multicolumn{2}{c|}{\multirow{3}[2]{*}{(1.5,1)}} & 5  & 0.7781 & 0.6158 & 0.0741 & 0.0766 & 0.0699 & 0.0080 & 0.1052 & 0.0293 & 0.1078 \\
			\multicolumn{2}{c|}{} & 10 & 0.5823 & 0.1947 & 0.0579 & 0.0687 & 0.0274 & 0.0075 & 0.0973 & 0.0144 & 0.0884 \\
			\multicolumn{2}{c|}{} & 15 & 0.5694 & 0.1114 & 0.0547 & 0.0645 & 0.0161 & 0.0072 & 0.0895 & 0.0095 & 0.0758 \\
			\midrule
			\multicolumn{2}{c|}{\multirow{3}[2]{*}{(1,1.5)}} & 5  & 0.6265 & 0.6602 & 0.0473 & 0.0669 & 0.0790 & 0.0054 & 0.1184 & 0.0205 & 0.1458 \\
			\multicolumn{2}{c|}{} & 10 & 0.5195 & 0.3003 & 0.0396 & 0.0498 & 0.0367 & 0.0051 & 0.1066 & 0.0109 & 0.1199 \\
			\multicolumn{2}{c|}{} & 15 & 0.4970 & 0.1991 & 0.0402 & 0.0460 & 0.0268 & 0.0046 & 0.0911 & 0.0079 & 0.0982 \\
			\midrule
			\multicolumn{2}{c|}{\multirow{3}[2]{*}{(1.5,1.5)}} & 5  & 0.7412 & 0.7977 & 0.0575 & 0.0874 & 0.0960 & 0.0070 & 0.0899 & 0.0235 & 0.1076 \\
			\multicolumn{2}{c|}{} & 10 & 0.6611 & 0.3901 & 0.0516 & 0.0687 & 0.0396 & 0.0060 & 0.0780 & 0.0106 & 0.0824 \\
			\multicolumn{2}{c|}{} & 15 & 0.5900 & 0.2081 & 0.0454 & 0.0706 & 0.0274 & 0.0060 & 0.0725 & 0.0079 & 0.0708 \\
			\bottomrule
		\end{tabular}%
		\label{tab:mcmc.rec}%
	\end{table}%
	
	\begin{table}[htbp]
		\centering
		\caption{Risks of Lindley Bayes estimates of unknown quantities based on order statistics ($t= 0.5$)}
		\begin{tabular}{cc|c|ccc|ccc|ccc}
			\toprule
			\multicolumn{2}{c|}{\multirow{2}[4]{*}{$(\alpha,\beta)$}} & \multirow{2}[4]{*}{$n$} & \multicolumn{3}{c|}{SELF} & \multicolumn{3}{c|}{LINEX} & \multicolumn{3}{c}{GELF } \\
			\cmidrule{4-12}    \multicolumn{2}{c|}{} &    & $\hat{\alpha}_{risk}$ & $\hat{\beta}_{risk}$ & $\hat{R(t)}_{risk}$ & $\hat{\alpha}_{risk}$ & $\hat{\beta}_{risk}$ & $\hat{R(t)}_{risk}$ & $\hat{\alpha}_{risk}$ & $\hat{\beta}_{risk}$ & \multicolumn{1}{c}{$\hat{R(t)}_{risk}$} \\
			\midrule
			\multicolumn{12}{c}{$(a_1,a_2,b_1,b_2)=(2,2,2,2)$} \\
			\midrule
			\multicolumn{2}{c|}{\multirow{3}[2]{*}{(1,1)}} & 5  & 0.0399 & 0.0642 & 0.0079 & 0.0052 & 0.0079 & 0.0010 & 0.0077 & 0.0103 & 0.0017 \\
			\multicolumn{2}{c|}{} & 10 & 0.0626 & 0.0643 & 0.0083 & 0.0078 & 0.0081 & 0.0011 & 0.0087 & 0.0078 & 0.0013 \\
			\multicolumn{2}{c|}{} & 15 & 0.0621 & 0.0449 & 0.0075 & 0.0075 & 0.0055 & 0.0009 & 0.0076 & 0.0053 & 0.0011 \\
			\midrule
			\multicolumn{2}{c|}{\multirow{3}[2]{*}{(1.5,1)}} & 5  & 0.2937 & 0.0861 & 0.0284 & 0.0404 & 0.0111 & 0.0032 & 0.0419 & 0.0150 & 0.0034 \\
			\multicolumn{2}{c|}{} & 10 & 0.0903 & 0.0572 & 0.0107 & 0.0105 & 0.0070 & 0.0014 & 0.0133 & 0.0067 & 0.0017 \\
			\multicolumn{2}{c|}{} & 15 & 0.0764 & 0.0424 & 0.0079 & 0.0097 & 0.0052 & 0.0010 & 0.0108 & 0.0050 & 0.0012 \\
			\midrule
			\multicolumn{2}{c|}{\multirow{3}[2]{*}{(1,1.5)}} & 5  & 0.0339 & 0.2035 & 0.0058 & 0.0046 & 0.0153 & 0.0006 & 0.0070 & 0.0172 & 0.0019 \\
			\multicolumn{2}{c|}{} & 10 & 0.0533 & 0.0656 & 0.0069 & 0.0065 & 0.0080 & 0.0008 & 0.0070 & 0.0088 & 0.0009 \\
			\multicolumn{2}{c|}{} & 15 & 0.0567 & 0.0538 & 0.0063 & 0.0068 & 0.0068 & 0.0008 & 0.0069 & 0.0071 & 0.0009 \\
			\midrule
			\multicolumn{2}{c|}{\multirow{3}[2]{*}{(1.5,1.5)}} & 5  & 0.3050 & 0.2407 & 0.0198 & 0.0424 & 0.0248 & 0.0023 & 0.0422 & 0.0313 & 0.0025 \\
			\multicolumn{2}{c|}{} & 10 & 0.1020 & 0.0778 & 0.0078 & 0.0096 & 0.0090 & 0.0010 & 0.0121 & 0.0093 & 0.0013 \\
			\multicolumn{2}{c|}{} & 15 & 0.0691 & 0.0535 & 0.0067 & 0.0089 & 0.0067 & 0.0009 & 0.0104 & 0.0071 & 0.0010 \\
			\midrule
			\multicolumn{12}{c}{$(a_1,a_2,b_1,b_2)=(0.05,0.05,0.05,0.05)$} \\
			\midrule
			\multicolumn{2}{c|}{\multirow{3}[2]{*}{(1,1)}} & 5  & 0.1728 & 0.1628 & 0.0347 & 0.021 & 0.0203 & 0.0044 & 0.0236 & 0.0188 & 0.0057 \\
			\multicolumn{2}{c|}{} & 10 & 0.1131 & 0.1015 & 0.0172 & 0.0142 & 0.013 & 0.0021 & 0.0149 & 0.0121 & 0.0025 \\
			\multicolumn{2}{c|}{} & 15 & 0.0851 & 0.063 & 0.0112 & 0.0103 & 0.0079 & 0.0014 & 0.0102 & 0.0075 & 0.0016 \\
			\midrule
			\multicolumn{2}{c|}{\multirow{3}[2]{*}{(1.5,1)}} & 5  & 0.2026 & 0.1493 & 0.0329 & 0.024 & 0.0197 & 0.0042 & 0.0302 & 0.0187 & 0.0051 \\
			\multicolumn{2}{c|}{} & 10 & 0.1116 & 0.0889 & 0.0158 & 0.0136 & 0.0114 & 0.002 & 0.0155 & 0.0107 & 0.0022 \\
			\multicolumn{2}{c|}{} & 15 & 0.0874 & 0.0581 & 0.0104 & 0.0108 & 0.0073 & 0.0013 & 0.0117 & 0.0069 & 0.0014 \\
			\midrule
			\multicolumn{2}{c|}{\multirow{3}[2]{*}{(1,1.5)}} & 5  & 0.1714 & 0.0996 & 0.0302 & 0.0215 & 0.0121 & 0.0038 & 0.0241 & 0.0142 & 0.005 \\
			\multicolumn{2}{c|}{} & 10 & 0.1181 & 0.0705 & 0.0164 & 0.0147 & 0.0088 & 0.002 & 0.0149 & 0.0092 & 0.0024 \\
			\multicolumn{2}{c|}{} & 15 & 0.0868 & 0.0597 & 0.0111 & 0.0108 & 0.0075 & 0.0014 & 0.0109 & 0.0077 & 0.0016 \\
			\midrule
			\multicolumn{2}{c|}{\multirow{3}[2]{*}{(1.5,1.5)}} & 5  & 0.1872 & 0.0975 & 0.0307 & 0.0224 & 0.0126 & 0.0039 & 0.0287 & 0.0149 & 0.005 \\
			\multicolumn{2}{c|}{} & 10 & 0.1157 & 0.0696 & 0.0168 & 0.0143 & 0.0087 & 0.0021 & 0.0163 & 0.0091 & 0.0024 \\
			\multicolumn{2}{c|}{} & 15 & 0.0879 & 0.0597 & 0.0111 & 0.0109 & 0.0073 & 0.0014 & 0.0119 & 0.0075 & 0.0015 \\
			\bottomrule
		\end{tabular}%
		\label{tab:lind.order}%
	\end{table}%
	\begin{table}[htbp]
		\centering
		\caption{Risks of T-K Bayes estimates of unknown quantities based on order statistics ($t= 0.5$)}
		\begin{tabular}{cc|r|ccc|ccc|ccc}
			\toprule
			\multicolumn{2}{c|}{\multirow{2}[4]{*}{$(\alpha,\beta)$}} & \multicolumn{1}{c|}{\multirow{2}[4]{*}{$n$}} & \multicolumn{3}{c|}{SELF} & \multicolumn{3}{c|}{LINEX} & \multicolumn{3}{c}{GELF } \\
			\cmidrule{4-12}    \multicolumn{2}{c|}{} &    & $\hat{\alpha}_{risk}$ & $\hat{\beta}_{risk}$ & $\hat{R(t)}_{risk}$ & $\hat{\alpha}_{risk}$ & $\hat{\beta}_{risk}$ & $\hat{R(t)}_{risk}$ & $\hat{\alpha}_{risk}$ & $\hat{\beta}_{risk}$ & \multicolumn{1}{c}{$\hat{R(t)}_{risk}$} \\
			\midrule
			\multicolumn{12}{c}{$(a_1,a_2,b_1,b_2)=(2,2,2,2)$} \\
			\midrule
			\multicolumn{2}{c|}{\multirow{3}[2]{*}{(1,1)}} & \multicolumn{1}{c|}{5} & 0.1565 & 0.1548 & 0.0141 & 0.0164 & 0.0178 & 0.0017 & 0.0083 & 0.0119 & 0.0132 \\
			\multicolumn{2}{c|}{} & \multicolumn{1}{c|}{10} & 0.1109 & 0.0850 & 0.0113 & 0.0131 & 0.0104 & 0.0014 & 0.0003 & 0.0073 & 0.0088 \\
			\multicolumn{2}{c|}{} & \multicolumn{1}{c|}{15} & 0.0868 & 0.0505 & 0.0092 & 0.0107 & 0.0062 & 0.0012 & 0.0009 & 0.0083 & 0.0062 \\
			\midrule
			\multicolumn{2}{c|}{\multirow{3}[2]{*}{(1.5,1)}} & \multicolumn{1}{c|}{5} & 0.1547 & 0.1471 & 0.0179 & 0.0168 & 0.0168 & 0.0022 & 0.0079 & 0.0033 & 0.0115 \\
			\multicolumn{2}{c|}{} & \multicolumn{1}{c|}{10} & 0.1452 & 0.0749 & 0.0118 & 0.0164 & 0.0090 & 0.0015 & 0.0000 & 0.0185 & 0.0057 \\
			\multicolumn{2}{c|}{} & \multicolumn{1}{c|}{15} & 0.1131 & 0.0484 & 0.0080 & 0.0132 & 0.0059 & 0.0010 & 0.0039 & 0.0030 & 0.0034 \\
			\midrule
			\multicolumn{2}{c|}{\multirow{3}[2]{*}{(1,1.5)}} & \multicolumn{1}{c|}{5} & 0.1245 & 0.1593 & 0.0110 & 0.0128 & 0.0170 & 0.0014 & 0.0017 & 0.0055 & 0.0114 \\
			\multicolumn{2}{c|}{} & \multicolumn{1}{c|}{10} & 0.1038 & 0.1304 & 0.0096 & 0.0121 & 0.0153 & 0.0012 & 0.0000 & 0.0115 & 0.0083 \\
			\multicolumn{2}{c|}{} & \multicolumn{1}{c|}{15} & 0.0745 & 0.0843 & 0.0073 & 0.0091 & 0.0102 & 0.0009 & 0.0113 & 0.0002 & 0.0058 \\
			\midrule
			\multicolumn{2}{c|}{\multirow{3}[2]{*}{(1.5,1.5)}} & \multicolumn{1}{c|}{5} & 0.1298 & 0.1371 & 0.0124 & 0.0145 & 0.0147 & 0.0015 & 0.0002 & 0.0169 & 0.0097 \\
			\multicolumn{2}{c|}{} & \multicolumn{1}{c|}{10} & 0.1310 & 0.1156 & 0.0095 & 0.0145 & 0.0136 & 0.0012 & 0.0018 & 0.0170 & 0.0058 \\
			\multicolumn{2}{c|}{} & \multicolumn{1}{c|}{15} & 0.1069 & 0.0785 & 0.0075 & 0.0125 & 0.0094 & 0.0009 & 0.0008 & 0.0006 & 0.0039 \\
			\midrule
			\multicolumn{12}{c}{$(a_1,a_2,b_1,b_2)=(0.05,0.05,0.05,0.05)$} \\
			\midrule
			\multicolumn{2}{c|}{\multirow{3}[2]{*}{(1,1)}} & 5  & 0.4364 & 0.3500 & 0.0246 & 0.0414 & 0.0419 & 0.0030 & 0.0002 & 0.0846 & 0.0224 \\
			\multicolumn{2}{c|}{} & 10 & 0.1841 & 0.1339 & 0.0144 & 0.0220 & 0.0173 & 0.0018 & 0.0091 & 0.0206 & 0.0119 \\
			\multicolumn{2}{c|}{} & 15 & 0.0893 & 0.0670 & 0.0101 & 0.0109 & 0.0084 & 0.0013 & 0.0100 & 0.0298 & 0.0074 \\
			\midrule
			\multicolumn{2}{c|}{\multirow{3}[2]{*}{(1.5,1)}} & 5  & 0.8246 & 0.3618 & 0.0247 & 0.0598 & 0.0436 & 0.0030 & 0.0454 & 0.0114 & 0.0170 \\
			\multicolumn{2}{c|}{} & 10 & 0.3987 & 0.1218 & 0.0149 & 0.0425 & 0.0151 & 0.0018 & 0.0113 & 0.0004 & 0.0069 \\
			\multicolumn{2}{c|}{} & 15 & 0.2270 & 0.0694 & 0.0095 & 0.0268 & 0.0086 & 0.0012 & 0.0108 & 0.0003 & 0.0038 \\
			\midrule
			\multicolumn{2}{c|}{\multirow{3}[2]{*}{(1,1.5)}} & 5  & 0.4085 & 0.7131 & 0.0226 & 0.0385 & 0.0823 & 0.0028 & 0.0093 & 0.0237 & 0.0408 \\
			\multicolumn{2}{c|}{} & 10 & 0.1834 & 0.2342 & 0.0141 & 0.0219 & 0.0291 & 0.0017 & 0.0001 & 0.0011 & 0.0158 \\
			\multicolumn{2}{c|}{} & 15 & 0.0997 & 0.1414 & 0.0096 & 0.0123 & 0.0176 & 0.0012 & 0.0098 & 0.0023 & 0.0088 \\
			\midrule
			\multicolumn{2}{c|}{\multirow{3}[2]{*}{(1.5,1.5)}} & 5  & 0.7049 & 0.5513 & 0.0242 & 0.0513 & 0.0608 & 0.0030 & 0.0843 & 0.0011 & 0.0216 \\
			\multicolumn{2}{c|}{} & 10 & 0.3847 & 0.2198 & 0.0135 & 0.0414 & 0.0265 & 0.0017 & 0.0043 & 0.0057 & 0.0083 \\
			\multicolumn{2}{c|}{} & 15 & 0.2394 & 0.1348 & 0.0103 & 0.0281 & 0.0166 & 0.0013 & 0.0014 & 0.0000 & 0.0053 \\
			\bottomrule
		\end{tabular}%
		\label{tab:tk.order}%
	\end{table}%
	\begin{table}[htbp]
		\centering
		\caption{Risks of MCMC Bayes estimates of unknown quantities based on order statistics ($t= 0.5$)}
		\begin{tabular}{cc|c|ccc|c|cc|ccc}
			\toprule
			\multicolumn{2}{c|}{\multirow{2}[4]{*}{$(\alpha,\beta)$}} & \multirow{2}[4]{*}{$n$} & \multicolumn{3}{c|}{SELF} & \multicolumn{3}{c|}{LINEX} & \multicolumn{3}{c}{GELF } \\
			\cmidrule{4-12}    \multicolumn{2}{c|}{} &    & $\hat{\alpha}_{risk}$ & $\hat{\beta}_{risk}$ & $\hat{R(t)}_{risk}$ & \multicolumn{1}{c}{$\hat{\alpha}_{risk}$} & $\hat{\beta}_{risk}$ & $\hat{R(t)}_{risk}$ & $\hat{\alpha}_{risk}$ & $\hat{\beta}_{risk}$ & \multicolumn{1}{c}{$\hat{R(t)}_{risk}$} \\
			\midrule
			\multicolumn{12}{c}{$(a_1,a_2,b_1,b_2)=(2,2,2,2)$} \\
			\midrule
			\multicolumn{2}{c|}{\multirow{3}[2]{*}{(1,1)}} & 5  & 0.1697 & 0.1478 & 0.0145 & \multicolumn{1}{c}{0.0188} & 0.0167 & 0.0019 & 0.0178 & 0.0105 & 0.0136 \\
			\multicolumn{2}{c|}{} & 10 & 0.0953 & 0.0894 & 0.0102 & \multicolumn{1}{c}{0.0108} & 0.0108 & 0.0012 & 0.0102 & 0.0071 & 0.0074 \\
			\multicolumn{2}{c|}{} & 15 & 0.0737 & 0.0511 & 0.0087 & \multicolumn{1}{c}{0.0087} & 0.0058 & 0.0010 & 0.0079 & 0.0045 & 0.0051 \\
			\midrule
			\multicolumn{2}{c|}{\multirow{3}[2]{*}{(1.5,1)}} & 5  & 0.1812 & 0.1342 & 0.0176 & \multicolumn{1}{c}{0.0228} & 0.0125 & 0.0026 & 0.0194 & 0.0097 & 0.0128 \\
			\multicolumn{2}{c|}{} & 10 & 0.1645 & 0.0758 & 0.0121 & \multicolumn{1}{c}{0.0156} & 0.0086 & 0.0013 & 0.0098 & 0.0064 & 0.0053 \\
			\multicolumn{2}{c|}{} & 15 & 0.1290 & 0.0478 & 0.0073 & \multicolumn{1}{c}{0.0130} & 0.0072 & 0.0011 & 0.0073 & 0.0052 & 0.0039 \\
			\midrule
			\multicolumn{2}{c|}{\multirow{3}[2]{*}{(1,1.5)}} & 5  & 0.1376 & 0.1662 & 0.0122 & \multicolumn{1}{c}{0.0137} & 0.0157 & 0.0014 & 0.0132 & 0.0094 & 0.0108 \\
			\multicolumn{2}{c|}{} & 10 & 0.0911 & 0.1104 & 0.0092 & \multicolumn{1}{c}{0.0122} & 0.0120 & 0.0012 & 0.0106 & 0.0053 & 0.0085 \\
			\multicolumn{2}{c|}{} & 15 & 0.0753 & 0.0783 & 0.0080 & \multicolumn{1}{c}{0.0077} & 0.0105 & 0.0009 & 0.0071 & 0.0046 & 0.0060 \\
			\midrule
			\multicolumn{2}{c|}{\multirow{3}[2]{*}{(1.5,1.5)}} & 5  & 0.1699 & 0.1259 & 0.0141 & \multicolumn{1}{c}{0.0192} & 0.0179 & 0.0017 & 0.0152 & 0.0100 & 0.0099 \\
			\multicolumn{2}{c|}{} & 10 & 0.1257 & 0.1164 & 0.0097 & \multicolumn{1}{c}{0.0147} & 0.0134 & 0.0012 & 0.0091 & 0.0062 & 0.0058 \\
			\multicolumn{2}{c|}{} & 15 & 0.0988 & 0.0765 & 0.0061 & \multicolumn{1}{c}{0.0119} & 0.0095 & 0.0010 & 0.0069 & 0.0039 & 0.0043 \\
			\midrule
			\multicolumn{12}{c}{$(a_1,a_2,b_1,b_2)=(0.05,0.05,0.05,0.05)$} \\
			\midrule
			\multicolumn{2}{c|}{\multirow{3}[2]{*}{(1,1)}} & 5  & 0.3344 & 0.4099 & 0.0273 & 0.0562 & 0.0549 & 0.0036 & 0.0462 & 0.0224 & 0.0412 \\
			\multicolumn{2}{c|}{} & 10 & 0.2050 & 0.1222 & 0.0177 & 0.0254 & 0.0201 & 0.0021 & 0.0195 & 0.0106 & 0.0148 \\
			\multicolumn{2}{c|}{} & 15 & 0.1101 & 0.0629 & 0.0110 & 0.0118 & 0.0087 & 0.0014 & 0.0117 & 0.0059 & 0.0087 \\
			\midrule
			\multicolumn{2}{c|}{\multirow{3}[2]{*}{(1.5,1)}} & 5  & 0.5720 & 0.3512 & 0.0252 & 0.0649 & 0.0412 & 0.0037 & 0.0300 & 0.0198 & 0.0211 \\
			\multicolumn{2}{c|}{} & 10 & 0.4530 & 0.1176 & 0.0161 & 0.0408 & 0.0137 & 0.0019 & 0.0149 & 0.0088 & 0.0070 \\
			\multicolumn{2}{c|}{} & 15 & 0.2642 & 0.0599 & 0.0111 & 0.0242 & 0.0071 & 0.0012 & 0.0093 & 0.0056 & 0.0038 \\
			\midrule
			\multicolumn{2}{c|}{\multirow{3}[2]{*}{(1,1.5)}} & 5  & 0.4867 & 0.5031 & 0.0253 & 0.0403 & 0.0676 & 0.0031 & 0.0403 & 0.0167 & 0.0466 \\
			\multicolumn{2}{c|}{} & 10 & 0.1473 & 0.2134 & 0.0137 & 0.0239 & 0.0366 & 0.0019 & 0.0185 & 0.0100 & 0.0188 \\
			\multicolumn{2}{c|}{} & 15 & 0.1315 & 0.1739 & 0.0120 & 0.0107 & 0.0158 & 0.0012 & 0.0099 & 0.0056 & 0.0088 \\
			\midrule
			\multicolumn{2}{c|}{\multirow{3}[2]{*}{(1.5,1.5)}} & 5  & 0.5585 & 0.5337 & 0.0258 & 0.0606 & 0.0552 & 0.0032 & 0.0266 & 0.0162 & 0.0238 \\
			\multicolumn{2}{c|}{} & 10 & 0.3194 & 0.2301 & 0.0138 & 0.0433 & 0.0271 & 0.0018 & 0.0148 & 0.0082 & 0.0086 \\
			\multicolumn{2}{c|}{} & 15 & 0.2483 & 0.1506 & 0.0117 & 0.0265 & 0.0170 & 0.0013 & 0.0098 & 0.0056 & 0.0055 \\
			\bottomrule
		\end{tabular}%
		\label{tab:mcmc.order}%
	\end{table}%
\section{Numerical Illustration: Cotton Production Data}
\begin{figure}[h]
	\centering
	\includegraphics[width=0.7\linewidth]{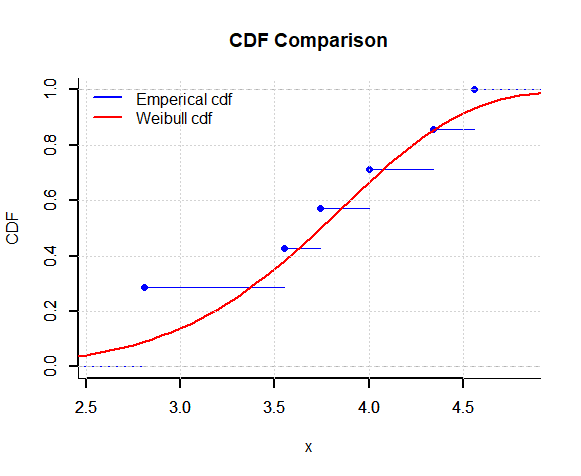}
	\caption{Plot of empirical and theoretical cdf}
	\label{fig:weibullfitting}
\end{figure}

To see the real world applicability of the developed results, we have taken the data of production of cotton in the United States of America. USA is among the top producers of cotton worldwide. U.S cotton production varies annually depending on weather trends, market preferences and governmental regulations. 
\begin{table}[htbp]
	\centering
	\caption{US cotton Production From 2013-2014 to 2019- 2020}
	\begin{tabular}{cc}
		\toprule
		\textbf{Year} & \textbf{Production}  \\
		\midrule
		2013-14 & 2.81  \\
		2014-15 & 3.55  \\
		2015-16  & 2.81 \\
		2016-17  & 3.74  \\
		2017-18 & 4.56  \\
		2018-19 & 4 \\
		2019-20 & 4.34  \\
		\bottomrule
	\end{tabular}%
	\label{tab:realdatacrime}%
\end{table}%
As it is mentioned clear in the Introduction (Section [\ref{secintro}]) that the transformation $Y=e^{-X}$ follows UW distribution when $X$ follows Weibull distribution. So the authors decided that first its best to see the support of the data with respect to Weibull distribution and then by taking the transformation we will check the fitting for the UW distribution. In this direction, we frame hypotheses as $H_o: F(x)=F_n(x)$ against $H_1:F(x)\ne F_n(x),$ where $F(x)$ represents cdf of Weibull distribution and $F_n(x)$ represents empirical distribution of data. Using Kolmogorov-Smirnov (KS) test, we find that $D = 0.1944$, p-value $= 0.954$, which supports that weibull is a good fit.This claim is being supported by the visualization given in Figure [\ref{fig:weibullfitting}]. To ensure that this distribution provides best fit among other distribution then we will commpare the fitting of the distribution with some well known distributions (See Table [\ref{fitting_compare}]). For this purpose, we are using the criteria of Akaike Information Criterion (AIC), Bayesian Information Criterion (BIC) and log likelihood. The estimates are obtained using maximum likelihood techniques.

\begin{table}[htbp]
	\centering
	\caption{Comparison of fitting results of data}
	\begin{tabular}{ccccc}
		\toprule
		\textbf{Distribution} & \textbf{Estimates}  & \textbf{-log likelihood} & \textbf{AIC} & \textbf{BIC} \\
		\midrule
		
		Weibull&	$\alpha= 6.89,~ \beta=3.95$ &	-6.643325&	17.28665	& 17.17847 \\
		Gamma &	$\alpha= 31.96,~ \beta=8.66$ & -6.866839  &		17.73368 &	17.6255 \\
		Normal &	$\mu= 3.68,~ \sigma=0.63$  &	-6.783007 &	17.56601 &	17.45783 \\
		Exponential	& $\alpha= 0.27$ &	-16.13396 &	34.26793 &	34.21384 \\
		
		\bottomrule
	\end{tabular}%
	\label{fitting_compare}%
\end{table}%

Now, we will show that the transformed data supports UW distribution. For this we apply KS test and it is found that for $\alpha=6.89 ,$ $\beta=7.67\times 10^{-5}$ the KS statistics is 0.1958 with $p-$ value 0.9525. As this article provides generalized nature of results for ordered random variables. Now we will present our results in the context of order statistics. From the data given in Table [\ref{tab:realdatacrime}], the order statistics can easily be found. Based on order statistics, the Bayes estimates (for $m=0$ and $k=1$) of the unknown parameters are calculated and reported in Table [\ref{tab:order_est}]. \par
Now, we show the application of lower record values in the context of the same application. For this purpose, we first extract lower record values from the transformed dataset. The obtained lower record values are 0.0602, 0.0287, 0.0238, 0.0183, 0.0130, 0.0105.  Based on record values, the Bayes estimates (for $m=-1$ and $k=1$) of the unknown parameters are calculated and reported in Table [\ref{tab:record_est}].

\begin{table}[htbp]
	\centering
	\caption{Bayes estimates of UW$(\alpha,\beta)$ for $c=0.5$, $t=0.5$ and Prior I: Order Statistics}
	\begin{tabular}{l|rrr|rrr|rrr}
		\toprule
		\multirow{2}[3]{*}{Method} &       & \multicolumn{1}{c}{SELF} &       &       & \multicolumn{1}{c}{LINEX} &       &       & \multicolumn{1}{c}{GELF} &  \\
		\cmidrule{2-10}          & \multicolumn{1}{c}{$\alpha$} & \multicolumn{1}{c}{$\beta$} & \multicolumn{1}{c|}{$R(t)$} & \multicolumn{1}{c}{$\alpha$} & \multicolumn{1}{c}{$\beta$} & \multicolumn{1}{c|}{$R(t)$} & \multicolumn{1}{c}{$\alpha$} & \multicolumn{1}{c}{$\beta$} & \multicolumn{1}{c}{$R(t)$} \\
		\midrule
		Lindley & 0.4876 & 1.2564 & 0.2134 & 0.4981 & 1.6754 & 0.2806 & 0.4129 & 1.6987 & 0.1659 \\
		T-K   & 0.5969 & 1.8678 & 0.2615 & 0.5354 & 1.8043 & 0.2513 & 0.4293 & 1.7724 & 0.1840 \\
		MCMC  & 0.2508 & 2.9533 & 0.1057 & 0.2183 & 2.7360 & 0.0989 & 0.0563 &  2.6828 & 0.0156 \\
		\bottomrule
	\end{tabular}%
	\label{tab:order_est}%
\end{table}
\begin{table}[htbp]
	\centering
	\caption{Bayes estimates of UW$(\alpha,\beta)$ for $c=0.5$, $t=0.5$ and Prior I: Lower Record Values}
	\begin{tabular}{l|rrr|rrr|rrr}
		\toprule
		\multirow{2}[3]{*}{Method} &       & \multicolumn{1}{c}{SELF} &       &       & \multicolumn{1}{c}{LINEX} &       &       & \multicolumn{1}{c}{GELF} &  \\
		\cmidrule{2-10}          & \multicolumn{1}{c}{$\alpha$} & \multicolumn{1}{c}{$\beta$} & \multicolumn{1}{c|}{$R(t)$} & \multicolumn{1}{c}{$\alpha$} & \multicolumn{1}{c}{$\beta$} & \multicolumn{1}{c|}{$R(t)$} & \multicolumn{1}{c}{$\alpha$} & \multicolumn{1}{c}{$\beta$} & \multicolumn{1}{c}{$R(t)$} \\
		\midrule
		Lindley & 0.1231 & 1.5871 & 0.0876 & 0.1023 & 1.6251 & 0.0871 & 0.1104 & 1.6326 & 0.0611 \\
		T-K   & 0.1461 & 1.8000 & 0.0797 & 0.1269 & 1.7078 & 0.0629 & 0.1004 & 1.6811 & 0.0482 \\
		MCMC  & 0.2340 & 2.9930 & 0.1049 & 0.2038 & 2.6531 & 0.0982 & 0.0302 &  2.5683 & 0.0074 \\
		\bottomrule
	\end{tabular}%
	\label{tab:record_est}%
\end{table}%
 \section*{Statements and Declarations}
\begin{itemize}
	\item \textbf{Conflict of Interest:} On behalf of all authors, the corresponding author states that there is no conflict of interest.
	\item \textbf{Funding Information:} There is no funding available.
	\item \textbf{Data availability:} The authors confirm that the data supporting the findings of this study are available in the article (See Table \ref{tab:realdatacrime}).
	\item \textbf{Ethics approval:} Not Applicable

\end{itemize}

\bibliographystyle{apa}
\bibliography{ref.bib}

\end{document}